\documentclass{aastex}
\usepackage{emulateapj5}
\usepackage{graphicx}

\newcommand{\bq}{\begin{equation}}
\newcommand{\eq}{\end{equation}}

\shorttitle{Photometric Redshifts for GOODS Galaxies}
\shortauthors{Mobasher {\it et al}}

\begin{document}

\title{Photometric Redshifts for Galaxies in the GOODS Southern Field}
\author{
B.\ Mobasher\altaffilmark{3,4},
R.\ Idzi\altaffilmark{5}, 
N.\ Ben\'\i tez\altaffilmark{5},
A.\ Cimatti\altaffilmark{6}, 
S.\ Cristiani,\altaffilmark{7}, 
E.\ Daddi\altaffilmark{8}, 
T.\ Dahlen\altaffilmark{3}, 
M.\ Dickinson\altaffilmark{3,5}, 
T.\ Erben\altaffilmark{9}, 
H.\ C.\ Ferguson\altaffilmark{3,5}, 
M.\ Giavalisco\altaffilmark{3},
N.\ A.\ Grogin\altaffilmark{5}, 
A.\ M.\ Koekemoer\altaffilmark{3}, 
M.\ Mignoli\altaffilmark{10},
L.\ A.\ Moustakas\altaffilmark{3}, 
M.\ Nonino\altaffilmark{7}, 
P.\ Rosati\altaffilmark{8},
M.\ Schirmer\altaffilmark{9,11}, 
D.\ Stern\altaffilmark{12}, 
E.\ Vanzella,\altaffilmark{8},
C.\ Wolf\altaffilmark{13}, 
G.\ Zamorani\altaffilmark{10}
}

\altaffiltext{1}{
Based on observations taken with the NASA/ESA Hubble
Space Telescope, which is operated by the Association of Universities
for Research in Astronomy, Inc.\ (AURA) under NASA contract
NAS5--26555. This work is supported by NASA through grant GO09583.01-96A.
}

\altaffiltext{2}{
Based on observations collected at the European Southern 
Observatory, Chile (ESO Programmes 
168.A-0485,  170.A-0788, 64.O-0643,  66.A-0572, 68.A-0544, 164.O-0561, 
169.A-0725, 267.A-5729 66.A-0451,  68.A-0375  164.O-0561, 267.A-5729, 
169.A-0725,  64.O-0621)
}

\altaffiltext{3}{Space Telescope Science Institute, 
3700 San Martin Drive, Baltimore MD 21218, USA}
\altaffiltext{4}{Also affiliated to the Space Sciences 
Department of the European Space Agency}
\altaffiltext{5}{Physics and Astronomy Department, Johns Hopkins University, Baltimore, MD 21218, USA}
\altaffiltext{6}{Instituto Nazionale di Astrofisica, Osservatorio Astrofisico 
di Arcetri, Largo E. Fermi 5, 50125 Firenze, Italy}
\altaffiltext{7}{Istituto Nazionale di Astrofisica, Osservatorio Astronomico 
   di Trieste, via G.B. Tiepolo 11, Trieste, I--34131, Italy}
\altaffiltext{8}{European Southern Observatory, Karl-Schwarzschild- Str 2, 
85748 Garching, Germany}
\altaffiltext{9}{Institut f\"ur Astrophysik und Extraterrestrische 
  Forschung, Universit\"at Bonn, Auf dem H\"ugel 71, Bonn, Germany}
\altaffiltext{10}{Instituto Nazionale di Astrofisica, Osservatorio Astrofisico 
di Bologna, via Ranzani 1, 40127 Bolonga, Italy}
\altaffiltext{11}{Max-Planck-Institut f\"ur Astrophysik, 
       Karl--Schwarzschild--Str. 1,
       D--85748 Garching bei M\"unchen, Germany}
\altaffiltext{12} {Jet Propulsion Laboratory, California Institute of 
Technology, Mail Stop 169-506, Pasadena, CA 91109}
\altaffiltext{13}{Department of Physics, University of Oxford, Keble Road,
       Oxford, OX1 3RH, UK}

\begin{abstract}
We use extensive multi-wavelength photometric data from the 
Great Observatories Origins Deep Survey to estimate photometric 
redshifts for a sample of 434 galaxies with spectroscopic 
redshifts in the Chandra Deep Field South.  Using the Bayesian 
method (Ben\'\i tez 2000), which 
incorporates redshift/magnitude priors, we estimate 
photometric redshifts for galaxies in the range $18 < R_{AB} < 25.5$, 
giving an {\it rms} scatter 
$\sigma ({\Delta z \over 1 + z_{spec}}) \leq 0.11$. The outlier fraction is
$< 10\%$, with the outlier-clipped {\it rms} being 0.047. 
We examine the accuracy of photometric redshifts for several, 
special sub--classes of objects.  The results for extremely red 
objects are more accurate than those for the sample as a whole, 
with $\sigma = 0.051$ and very few outliers (3\%).  Photometric redshifts 
for active galaxies, identified from their X-ray 
emission,  have a dispersion of $\sigma =  0.104$, with 10\% outlier 
fraction, similar to that for normal galaxies. 
Employing a redshift/magnitude prior in this process seems to be 
crucial in improving the agreement between photometric and 
spectroscopic redshifts.

\end{abstract}


\section{INTRODUCTION}

The successful installation of the Advanced Camera for Surveys (ACS) 
on-board the Hubble Space Telescope (HST), and the promise of new 
observatories like the Space Infrared Telescope Facility (SIRTF), 
have opened new opportunities for multiwavelength surveys of the
distant universe.  However, most of the galaxies detected in these 
surveys are too faint for spectroscopic observations.  The photometric 
redshift technique is the only practical way to estimate distances 
to these galaxies, needed for statistical studies of their properties. 
The technique has the advantage of providing redshifts for large 
samples of faint galaxies with a relatively modest investment of 
observing time, but has the disadvantage of having coarse wavelength 
resolution ($\sim$ 1000 \AA) compared to a much higher resolution obtained 
spectroscopically ($\sim 10$ \AA).  

The Great Observatories Origins Deep Survey (GOODS) is designed to reach 
unprecedented depths in two fields, the Chandra Deep Field South (CDF-S) 
and Hubble Deep Field North (HDF-N), using HST, Chandra, and SIRTF.  
The majority of the objects detected in GOODS are fainter than the practical 
limit for spectroscopy.  Accurate photometric redshifts are therefore 
essential, in order to estimate distances to galaxies hosting type Ia 
supernovae, to convert observable properties (size, color, magnitude) to 
those in the rest-frame, and to explore many other aspects of galaxy 
evolution.  This {\em letter} provides details of the procedure used to 
estimate redshifts to galaxies in the GOODS CDF--S, and to assess 
the accuracy with which these are measured, using a sample of galaxies 
with spectroscopic redshifts.   We also test the accuracy of photometric 
redshifts for special subsamples of objects that have particular scientific 
interest within the GOODS data set: extremely red objects (EROs) and active 
galactic nuclei (AGNs).

% using ACS/HST ($z\sim 28$),
% Chandra ($S_{0.5-8 Kev} \sim 10^{-16} $ erg/sec/cm$^2$),
% IRAC/SIRTF ($S_{8 \mu m} \sim 1.2\ \mu Jy$)
% and MIPS/SIRTF ($S_{22\mu m}\sim 22\ \mu Jy$).

\section{The Sample}

\subsection{Photometric Observations}

We limit our present analysis to the GOODS CDF-S field, where the most 
complete, multiwavelength photometric data are currently available.  
The imaging data and photometric catalogs are described in Giavalisco 
et al.\ (2003).  Here, we have used ground-based optical ($U'UBVRI$) 
and near-infrared ($JHK_s$) data from ESO facilities 
(2.2m-WFI, VLT-FORS1, NTT-SOFI, VLT-ISAAC), as well as
from HST/ACS ($B_{435}V_{606}i_{775}z_{850}$).  Most of the imaging data 
cover the whole GOODS field of view, but a few (ISAAC $JHKs$, and 
FORS1 $RI$) do not.   In total, there are as many as 18 independent 
photometric measurements for each galaxy.  We matched the point spread 
functions (PSFs) of the images, including the ACS data, as described 
in Giavalisco et al.\ (2003).  An $R$-band selected catalog was created 
with SExtractor (Bertin \& Arnouts 1996).  The photometry used for the 
photometric redshifts was measured through matched, 3 arcsec diameter 
apertures in all bands. 

\subsection{Spectroscopic Observations}

The spectroscopic observations of CDF-S galaxies are from two 
separate samples:

\noindent {\bf K20:}  
A portion of this $K$--selected survey ($K_s({\rm Vega}) \leq 20$; 
$K_s({\rm AB}) \leq 21.85$; Cimatti et al 2002a, 2002b)   
lies within the CDF-S, and has reliable spectroscopic redshifts for 
271 objects, with a median redshift $\langle z \rangle=0.85$. Using
their spectra, the galaxies were classified as normal (i.e., emission and 
absorption line systems) or AGNs. 

\noindent {\bf GOODS/FORS2:}
The GOODS team is carrying out an extensive spectroscopic campaign
in the CDF-S as an ESO Large Programme (C.\ Cesarsky, PI).  Here, we 
use data from the first observing season, obtained with the VLT-FORS2 
spectrograph using the 300I grism ($R \approx 860$). The primary targets 
were faint ($z_{850} < 24.5$), red ($i_{775} - z_{850} > 0.6$) galaxies 
selected from the ACS imaging.  A heterogeneous sample of other galaxies 
was used to fill out the multislit masks.  We use 163 redshifts that were 
measured with good confidence, with a median redshift 
$\langle z \rangle = 1.05$.

\section{Photometric Redshift Technique}

We matched the spectroscopic and photometric catalogs, and used
the multicolor data to estimate photometric redshifts for galaxies
in the spectroscopic sample.   The comparison between photometric
and spectroscopic redshifts then gives an estimate of the photometric
redshift accuracy.  Because the two spectroscopic samples are selected 
differently, we have analyzed them separately and compare the results
to explore any biases induced by color/magnitude selection criteria.

We have performed extensive experiments with two variant methods of 
photometric redshift estimation: spectral template $\chi^2$ minimization 
(Puschell et al. 1982), and the Bayesian method of Ben\'\i tez (2000). 
In the $\chi^2$ technique, 
the photometric redshift of a galaxy is estimated by comparing its 
multi-band photometry with spectral energy distribution (SED) templates 
of galaxies with known types, shifted in redshift space, and integrated 
through the bandpass throughput functions.  For each template, at each 
redshift, the $\chi^2$ statistic is estimated, with the best--fitting
redshift and spectral type found from the minimum $\chi^2$ value 
assigned to the galaxy. 

The Bayesian approach considers the distribution $p(z|C,m)$, 
i.e. the redshift probability given not only the observed colors 
of a galaxy, $C$, 
but also its magnitude, $m$. This can be written as (Ben\'\i tez 2000):
$$
p(z|C,m) \propto \Sigma_T\ p(z,T|m) p(C|z,T)
$$

The first term on the right is the redshift/magnitude prior, 
which contains the probability of a galaxy having 
redshift $z$ and spectral type $T$, given its magnitude $m$, while
the second term is the redshift/type likelihood, 
$p(C|z,T)\propto exp(-\chi(z,T)^2/2.)$.

Note that using a prior of this kind is not equivalent to assuming 
a particular luminosity function; the prior describes 
the expected redshift distribution for galaxies of a certain magnitude, 
but does not include information about the galaxy magnitude distribution. 
The shape of the prior is estimated empirically from the HDF-N observations, 
as described in Ben\'\i tez (2000). The best estimate 
of the redshift, $z_B$, is then defined as the maximum of $p(z|C,m)$.  

  We compared results from several different software implementations of 
the $\chi^2$ method, as well as different choices of the SED templates
(empirical or synthetic), and found them to be in generally good agreement 
with one another and with the Bayesian 
Photometric Redshift (BPZ) results up to $z \sim 1$.  However, 
at higher redshifts, the BPZ approach gives significantly better 
performance, as described in \S4, below.  We have used the  
BPZ software \footnote{BPZ is available from 
http://acs.pha.jhu.edu/$\sim$txitxo/, Ben\'\i tez 2000} which offers both simple $\chi^2$ 
minimization, as well as the Bayesian estimate. 
We use template libraries, consisting of 
E, Sbc, Scd and Im SEDs from Coleman, Wu \& Weedman (1980) 
and two starburst SB2 and SB3 templates from Kinney et al.\ (1996). 
A two point interpolation between each pair of templates in the 
color--redshift space is performed, significantly improving 
the redshift resolution.
  
The BPZ uses a quality indicator, the 
Bayesian ODDS, which can be efficiently used to discard those objects 
with unreliable photo-z's. The ODDS is defined as the integral of 
the probability distribution 
$p(z|C,m)$ within a $0.27(1+z_B)$ interval centered on $z_B$. 
The ODDS would thus be 0.95 for a Gaussian $P(z|C,m)$ with 
width $\sigma = 0.067(1+z_B)$, 
which is the empirically measured accuracy of BPZ in the HDF-N. 

\section{Results}

\subsection{Overall Photometric Redshift Performance} 

We quantify the reliability of the photometric redshifts 
by measuring the fractional error for each galaxy, 
$\Delta \equiv {z_{phot} - z_{spec}\over 1 + z_{spec}}$.
We examine the median error, $\langle \Delta \rangle$, 
the {\it rms} scatter $\sigma(\Delta)$, and the rate of ``catastrophic'' 
outliers, $\eta$, defined as the fraction of the full sample 
that has $|\Delta| > 0.2$.

Table 1 presents the comparison between photometric and 
spectroscopic redshifts for both the K20 and FORS2 samples, 
using the complete photometric data set available for each object,
and for different combinations of magnitude limits and the ODDS 
parameter. Overall, the BPZ method yields $\sigma (\Delta) \approx 0.11$
and 0.07 for the FORS2 and K20 samples, respectively.  At fainter 
magnitudes ($R_{AB} > 25$), the BPZ method gives smaller {\it rms} 
scatter than the conventional $\chi^2$ minimization technique with 
no priors.  The BPZ results are presented in Figure 1.  The 
agreement between the photometric and spectroscopic redshifts 
is excellent, with median offset $\langle \Delta \rangle = -0.01$. 
There is no substantial trend in $\langle \Delta \rangle$ with 
redshift, except perhaps for a slight tendency for the photometric 
redshifts to be underestimated at $z > 1.3$. The {\it rms} values are all
based on a total of 433 galaxies in K20 (270) and FORS2 (163), with
the object at $z=2.8$ excluded (see 4.2b). 

For redshifts estimated with Bayesian priors, we find that the
fraction of catastrophic outliers, $\eta$, ranges from 0.024 to 0.10, 
depending on the subsample considered. As described in Ben\'\i tez (2000), 
the value of the ODDS parameter is a reasonable indicator of the reliability 
of the photometric redshift, galaxies with larger ODDS values 
have a smaller rate of outliers.  We have measured $\sigma (\Delta)$ 
using only galaxies with $|\Delta | < 0.2$ (i.e., excluding the
outliers), and find 0.049 (FORS2) and 0.046 (K20) from the BPZ method. 
The outlier--clipped {\it rms} is similar when using the conventional 
$\chi^2$ method without priors, but the failure rate, $\eta$, from
this method is much larger.

We have estimated redshifts using different combinations of 
the available photometric data in CDF-S, and find no significant 
differences in the $\sigma (\Delta)$ values. In particular, 
we have compared the performance with and without using the ACS 
photometry, and found no significant difference in the overall
result. The deep ACS photometry, however, may well be important at fainter 
magnitudes, beyond the limits of the spectroscopic sample.  
18\% of the FORS2 galaxies, and 13\% of the K20 galaxies, are 
undetected ($< 3\sigma$) in the relatively shallow WFI $U$ or $U'$ 
images.   However, we find no significant difference in $\Delta$ 
values for galaxies with and without $U/U'$ detections down to the 
spectroscopic magnitude limit (crosses in Figure 1).  We have also 
estimated photometric redshifts using the WFI and SOFI data alone, 
which cover a wider CDF--S area, including areas not imaged by ACS.
We measure $\sigma (\Delta) = 0.11$ (for $R_{AB} < 25$).

Figure 2 plots the redshift errors $\Delta$ versus galaxy
magnitudes in the $R$ and $K_s$ bands.  There is no 
strong trend in $\sigma(\Delta)$ with magnitude, except at the 
faintest $K_s$--band limits, $K_s \gtrsim 22.5$.  The outlier 
fraction $\eta$ increases from 0.03 ($K_s < 20$) to 0.05 ($20 < K_s < 22$) 
to 0.11 ($22 < K_s < 24$).  The photometric redshifts are also, 
on average, slightly underestimated for $K_s > 22$.  At these faint 
magnitudes, many of the galaxies are only poorly detected, if at all, 
in the wide--field (but relatively shallow) SOFI $JHK_s$ images, 
causing an increase in redshift errors. 

Results from Table 1 indicate that the photometric redshift performance
is correlated with the ODDS parameter, and is better for higher ODDS 
values (ODDS $>$ 0.99).  It is important to remember that the galaxies 
in the spectroscopic sample are relatively bright compared to the large 
majority of objects in the GOODS fields.  The ODDS parameter, perhaps 
combined with other indicators, (e.g., the number of passbands in which
the galaxy is significantly detected; availability of $U$-band and near-IR 
data; the width of the redshift probability distribution), offers a useful 
metric for the likely reliability of photometric redshifts at magnitudes 
fainter than the limit of our spectroscopic test sample.

\begin{table*}
\tablewidth{0pc}
\tabletypesize{\scriptsize}
% \caption{Comparison between photometric and spectroscopic redshifts for 
% the FORS2 and K20 galaxies in the 
% GOODS CDF-S using wfi+fors1+sofi+isaac+acs data\vspace{12pt}}
\caption{Comparison between photometric and spectroscopic redshifts}
\begin{tabular}{llllllllllll}
                      &      & \multicolumn{5}{c}{FORS2} &  \multicolumn{5}{c}{K20} \\
  $R_{lim}$ & ODDS & $\sigma (\Delta)$ &$\eta$ & $\sigma (\Delta)$ &$\eta$   & $n$ &  $\sigma (\Delta)$ &$\eta$&  $\sigma (\Delta)$  
 & $\eta$ &$n$\\
                      &      & \multicolumn{2}{c}{BPZ} &  \multicolumn{2}{c}{$\chi^2$} &  & \multicolumn{2}{c}{BPZ} &  \multicolumn{2}{c}{$\chi^2$} & \\
                       &      &           &           &   &           &           & &&& &\\
            &      &           & &          & &  &        &   &          & &\\ 

% \begin{deluxetable}{llllllllllll}
% \tablewidth{0pc}
% \tabletypesize{\scriptsize}
% \tablecaption{Comparison between photometric and spectroscopic redshifts}
% \tablehead{
%  & & \multicolumn{5}{c}{FORS2} & \multicolumn{5}{c}{K20} \\
% $R_{lim}$ & ODDS & $\sigma (\Delta)$ & $\eta$ & $\sigma (\Delta)$ &
% $\eta$ & $n$ &  $\sigma (\Delta)$ & $\eta$ &  $\sigma (\Delta)$ & $\eta$ & $n$ \\
% & & BPZ & & $\chi^2$ & & & BPZ &  & $\chi^2$ & & 
% }
% \startdata

     26.00 &     all&   0.107 & 0.098&  0.220& 0.11& 163&   0.072&0.048&  0.140&0.059&  270\\
     25.00 &     all&   0.114 & 0.099& 0.110& 0.110 &141& 0.072 & 0.046 & 0.142 & 0.058 & 261 \\ 
                   24.50 &     all  &  0.111 &0.100&  0.113&0.100&  111&   0.073&0.049&  0.146&0.049&  246\\   
                    25.00   &  $= 1.00$ &  0.082 &0.038&  0.083&0.038&  78 &   0.065&0.024&  0.068&0.030&  165\\
                    25.00    & $> 0.99$ &  0.095 &0.062&  0.098&0.062&  113&   0.064&0.028&  0.067&0.028&  246\\
                    25.00&     $> 0.95$ &  0.100 &0.073&  0.102&0.073&  124&   0.070&0.039&  0.073&0.043&  255\\
           &                &           &           &   &           &           & & & & & \\

% \enddata
% \end{deluxetable}

\end{tabular}
\end{table*}

\subsection{Special Galaxy Populations}

One of the main aims of the GOODS project is to identify and study 
different populations of galaxies, such as EROs and AGNs. 
In this section we carry out an analysis of the reliability of 
photometric redshifts to these objects.

\noindent {\bf a). EROs}--- The GOODS ERO sample is selected 
to have $(R-K)_{AB} > 3.35$ and is complete to $K_{AB} < 22$ mag.
A total of 66 EROs have spectroscopic redshifts in the combined 
FORS2 (36) and K20 (30) samples.  Figure 3a compares the photometric
and spectroscopic redshifts for EROs.  There is an excellent agreement,
with $\sigma (\Delta) = 0.051$, and is equally good for objects classified 
as absorption and emission line systems.  Furthermore, we find a very small
fraction of outliers ($\eta =3\%$). This performance is significantly 
{\it better} than that for the galaxy sample as a whole.  This may not 
be surprising:  in general, red galaxies have stronger features in their 
broad band spectral energy distributions (breaks, curvature) than do blue 
ones.  However, it is a very helpful result, because EROs are among the 
most difficult galaxies for spectroscopic observations.

The ERO population is known to consist of high redshift ($z\sim 1$) 
ellipticals and dusty starbursts (Cimatti et al 2002c).  However,
the starburst galaxy templates from Kinney et al (1996) that are
used for photometric redshift estimation are not significantly reddened, 
and certainly do not match the colors of EROs.  Photometric redshifts 
for the majority of the EROs, therefore, are derived from the elliptical 
galaxy spectral template, regardless of the true nature of the galaxies.  
Moustakas et al.\ (2003) find that $\sim 40\%$ of the GOODS EROs are 
morphologically early--type galaxies;  the rest are either disk galaxies 
or irregular systems.  However, they also find that the broad band 
SEDs of the EROs in different morphological 
subclasses are virtually indistinguishable.  Thus, we expect comparably 
good photometric redshift estimates for most EROs, regardless of their 
intrinsic nature. 

\noindent {\bf b). AGNs}--- 
48 galaxies in the spectroscopic K20 (31) and FORS2 (17) samples 
have X-ray detection (Alexander et al 2003). 
Figure 3b compares photometric and spectroscopic 
redshifts for these X-ray sources. The {\it rms} scatter, 
excluding the object at $z=2.8$, which is confirmed to be a QSO, 
is $\sigma (\Delta) = 0.104$, with an oulier fraction of $\eta = 0.11$. 
Excluding the five outliers reduces the {\it rms} scatter to 
$\sigma (\Delta) = 0.042$. 
The majority of the X-ray sources at higher redshifts here are AGNs although
some are likely to be X-ray starbursts. Six of these sources are
spectroscopically
confirmed as AGNs (crosses in Figure 3b). The scatter and the outlier fraction
for the X-ray sources (i.e. AGNs), is similar to that for normal galaxies 
in Figure 1. 
We tried using an independent set of observed AGN SEDs 
as templates, but did not significantly improve $\sigma(\Delta)$.
Using the conventional $\chi^2$ method without priors increases 
$\sigma (\Delta)$ value to 0.43.

\section{Conclusions}

Combining multi-waveband ground-based and ACS photometric data
(with up to 18 photometric passbands per object), we have estimated 
photometric redshifts for 433 galaxies with spectroscopic redshifts
in the GOODS CDF--S field, using the photometric redshift method 
of Ben\'\i tez 2000 which incorporates Bayesian magnitude/redshift priors.  
We find an 
overall performance $\sigma (\Delta) = 0.11$, with better performance 
for some subsamples.  The fraction of catastrophic redshift outliers
is $<$10\%, and is substantially smaller for galaxies with high 
values of the Bayesian ODDS parameter.  We see no strong trend 
in the performance of the photometric redshifts versus magnitudes,
except for an increase in the outlier rate.  Employing a redshift/magnitude
prior in this process seems to be crucial in reducing the scatter 
between photometric and spectroscopic redshifts.
 
We have applied the method to two sub--samples of galaxies
of particular interest:  EROs and AGNs.  The results for EROs are 
more accurate ($\sigma(\Delta) = 0.051$) than for faint galaxies 
in general.  They are somewhat less accurate for the X-ray sources 
(i.e. AGNs)- ($\sigma (\Delta)=0.104$), but good enough to be useful for 
many applications.  

%\acknowledgements

\clearpage

\begin{figure}
\epsscale{1}
\plotone{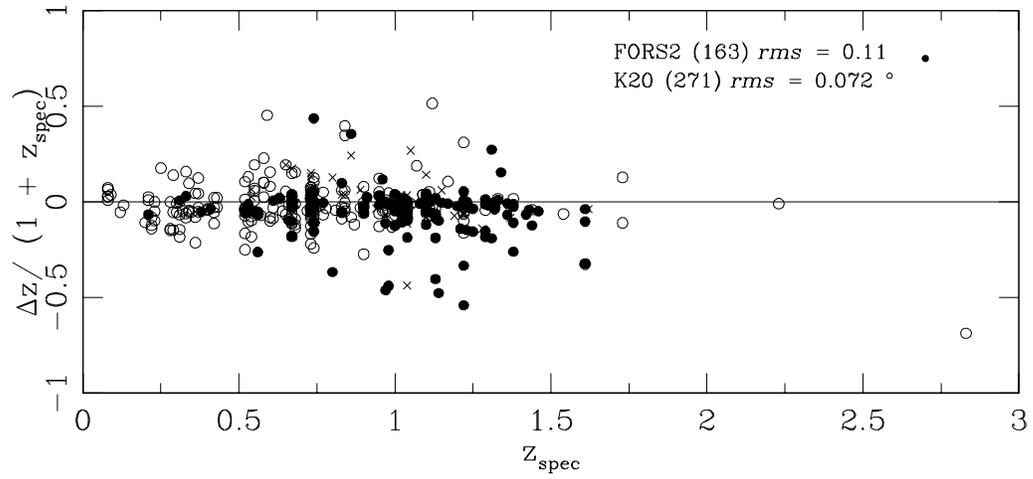}
\caption{Comparison between photometric and spectroscopic redshifts
for the K20 (open circles) and FORS (filled circles) samples. 
Galaxies undetected (very faint) in $U$ and $U'$ bands in both samples
are plotted by crosses. 
\label{fig1}} 
\end{figure}

\begin{figure}
\epsscale{1}
\plotone{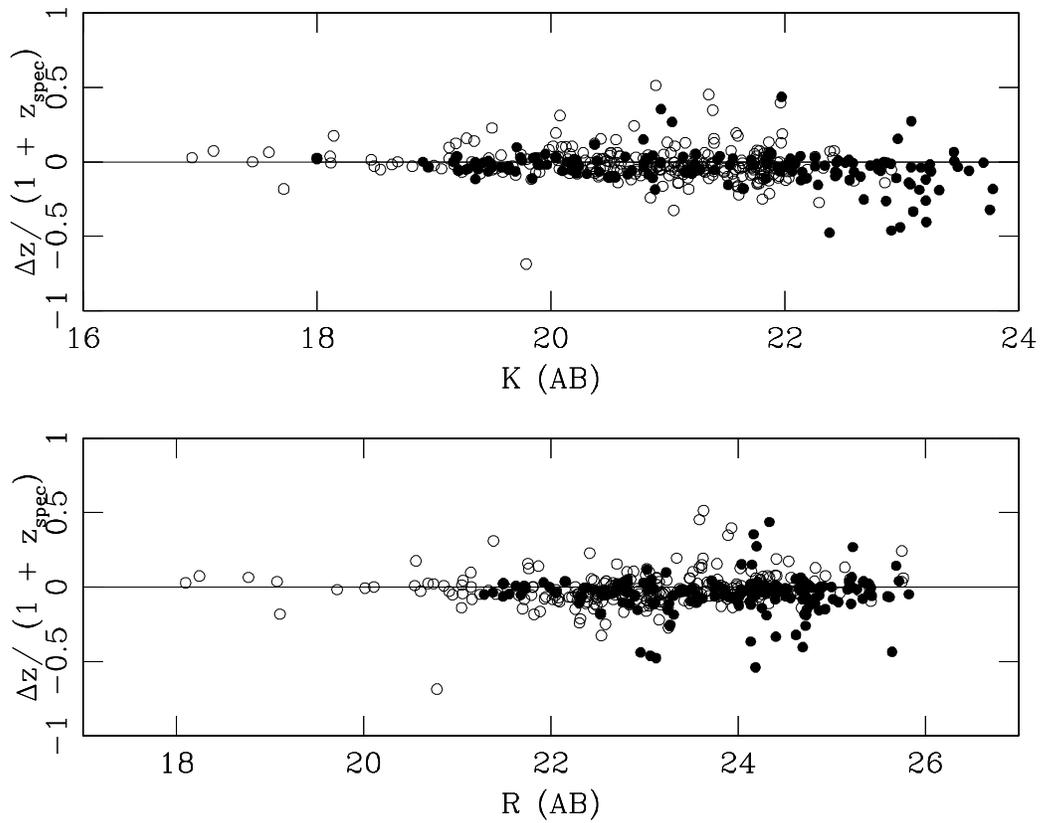}
\caption{Dependence of the accuracy of photometric redshifts to
brightness of galaxies in $R_{AB}$ and $K_{s_{AB}}$. 
Objects from both the K20 (open circles) and FORS (filled circles) samples
are included.
\label{fig2}} 
\end{figure}

\begin{figure}
\epsscale{1}
\plotone{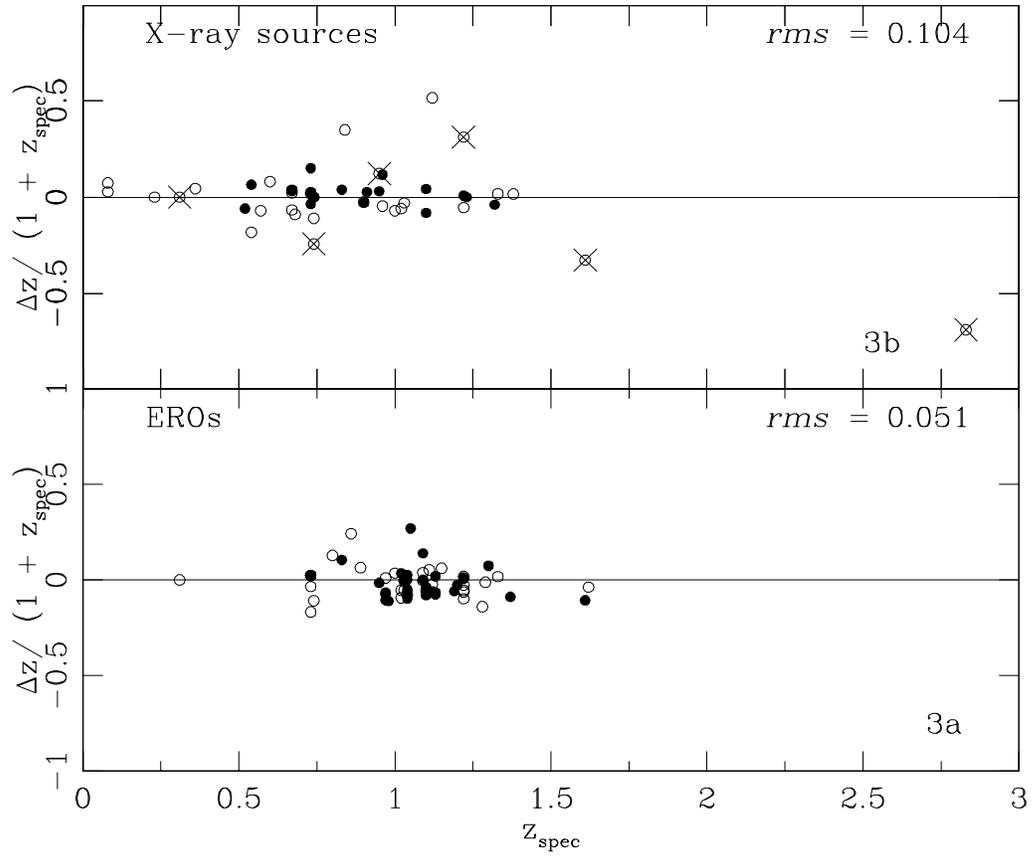}
\caption{Comparison between photometric and spectroscopic redshifts
for (3a). EROs from K20 (open circles) and FORS (filled circles). 
(3b). X-ray sources (i.e. AGNs) from K20 (open circles) and FORS2 
(filled circles), with spectroscopically confirmed AGNs (crosses). 
\label{fig3}} 
\end{figure}

\end{document}